\documentclass[aps,prb,showpacs,reprint, superscriptaddress]{revtex4-1}

\usepackage{graphicx}
\usepackage{color}
\usepackage{amsmath}
\usepackage{bbm}
\usepackage{amssymb}
 
\usepackage[T1]{fontenc}

\input epsf

\begin{document}

\title{Universal properties of high temperature superconductors from real space pairing: t-J-U model and its quantitative comparison with experiment}

\author{J{\'o}zef Spa{\l}ek}
\email{ufspalek@if.uj.edu.pl}
\affiliation{Marian Smoluchowski Institute of Physics, 
Jagiellonian University, ul. \L ojasiewicza 11,
30-348 Krak\'ow, Poland}

\author{Micha{\l} Zegrodnik}
\email{michal.zegrodnik@agh.edu.pl}
\affiliation{Academic Centre for Materials and Nanotechnology, AGH University of Science and Technology, Al. Mickiewicza 30, 30-059 Krakow,
Poland}

\author{Jan Kaczmarczyk}
\affiliation{Institute of Science and Technology Austria, Am Campus 1, A-34
00 Klosterneuburg, Austria}

\begin{abstract}
Selected universal experimental properties of high temperature superconducting (HTS) cuprates have been singled out in the last decade. One of the pivotal challenges in this field is the designation of a consistent interpretation framework within which we can describe quantitatively the universal features of those systems. Here we analyze in a detailed manner the principal experimental data and compare them quantitatively with the approach based on a single band of strongly correlated electrons supplemented with strong antiferromagnetic (super)exchange interaction (the so-called $t$-$J$-$U$ model). The model rationale is provided by estimating its macroscopic parameters on the basis of the 3-band approach for the Cu-O plane. We use our original full Gutzwiller-wave-function solution by going beyond the renormalized mean field theory (RMFT) in a systematic manner. Our approach reproduces very well the observed hole doping ($\delta$) dependence of the kinetic-energy gain in the superconducting phase, one of 
the principal non-Bardeen-Cooper-Schrieffer features of the cuprates. The calculated Fermi velocity in the nodal direction is practically $\delta$-independent and its universal 
value agrees very well with that determined experimentally. Also, a weak doping dependence of the Fermi wave-vector leads to an almost constant value of the effective mass in a pure superconducting phase which is both observed in the experiment and reproduced within our approach. An assessment of the currently used models is carried out and the results of the canonical RMFT as a zeroth-order solution are provided for comparison to illustrate the necessity of introduced higher order contributions. 
\end{abstract}


\maketitle

\section{Introduction}

High temperature superconductivity (HTS) in the quasi-two-dimensional cuprates is regarded as a fundamental phenomenon because of a number of reasons\cite{Bednorz1986,Schrieffer2007,Ogata2008,Uchida2015}. We name just a few here. First, by doping the system with holes (cf. Fig. \ref{fig:0}), a series of quantum phase transitions appears, starting from an antiferromagnetic Mott-Hubbard insulator (AFI) for the doping $\delta\lesssim0.02$ \cite{Bozovic2003}, through the HTS phase (often mixed with other phases up to an almost optimal doping\cite{Keimer2015}), to the normal Fermi-liquid-like state for $\delta\gtrsim0.3$ \cite{Kaminski2003}. On microscopic level, in the high-$\delta$ regime HTS disappears, most likely by a pair-correlation dilution in real space concomitant with an increased single-particle hopping via hole states. Second, when the doping is low, the pairing also weakens and the non-Bardeen-Cooper-Schrieffer (non-BCS) character of HTS shows up \cite{Bednorz1986,Deutscher2005,Carbonne2006,
Molegraaf2002,
Gianetti2011} in 
the form of the 
kinetic energy gain at the transition, which exemplifies the fact that the electronic-correlation effects are the strongest there. Third, the 
electronic spectrum in the nodal direction ($k_x=k_y$ in the two-dimensional Brillouin zone) exhibits a universal character\cite{Zhou2003,Shrakopi2008}. Namely, in spite of changing the carrier concentration by doping (which normally should lead to the corresponding changes in the Fermi-surface volume), the value of the Fermi velocity in the nodal direction remains almost unchanged in the whole doping range, where various phases such as superconducting, magnetically or charge-ordered can appear. Fourth, the effective mass in the nodal direction also exhibits a universal behavior\cite{Padilla2005}. These crucial features are regarded as representative to all the cuprate superconductors and our principal aim here is to address these and related properties in a fully quantitative manner.

It is accepted that the coper-oxide (CuO$_2$) planes which appear in the crystal structure of HTS are instrumental for achieving a stable paired phase\cite{Schrieffer2007,Ogata2008,Uchida2015,Bozovic2003}. That is why the theoretical analysis of the cuprates often limits to models which describe a single CuO$_2$\cite{Uchida2015} plane. Within such an approach one can eliminate the oxygen degrees-of-freedom via the Zhang-Rice singlet hypothesis \cite{Zhang1990,Zhang1988} or by perturbation expansion\cite{Zaanen1990,Eskes1993,Jefferson1992,Feiner1996,Avella2013,Emery1988,Valkov}, which leads to an effective single-band model of correlated $3d$ electrons due to Cu atoms on a square lattice (cf. Fig. 1). One of the canonical models for the description within the paradigm of strong correlations is the t-J model,\cite{Chao1976,Anderson1988,Spalek1988,Zhang1988,Spalek2007} in which HTS appears in the range $0<\delta\lesssim 0.4$ 
in a natural manner already within the \textit{renormalized mean field theory} RMFT\cite{Zhang1988_2,Edegger2007}, also with the so-called \textit{statistical consistency constraints} included explicitly (SGA method \cite{Jedrak2011,Spalek2015}). The RMFT approach in the SGA version can be related directly to the slave-boson method \cite{Comjai2007,Kotliar1988,Ruckenstein1987} (for review see Refs. \onlinecite{Lee2006,Jedrak2011_2}). However, within the Gutzwiller-type approach no extra Bose fields are required, as the interelectronic correlations are evaluated directly. Another model which is used in the theoretical analysis of HTS is the single-band Hubbard model which however requires more sophisticated calculation methods than RMFT to obtain the paired phase stability. The difference between the approach based on the $t$-$J$ model and the one that starts from the Hubbard model is that in the former case the intersite pairing corrections are included already via kinetic exchange whereas in the latter 
model one has to 
introduce them by including correlations beyond RMFT, as discussed also below. 

Recently, the full Gutzwiller wave-function (GWF) solution for the superconducting state for both the $t$-$J$ \cite{Kaczmarczyk2014} and the Hubbard\cite{Kaczmarczyk2013} models have been reported, in which RMFT (in the SGA form), 
appears as the zeroth-
order approximation to the full solution. Within 
this approach one can track down the evolution of the results, by using the so-called diagrammatic expansion method\cite{Bunemann2012, Gebhard1990} (DE-GWF), starting from the mean-field theory as the zeroth order result and proceed with incorporating systematically the nonlocal 
correlations of increased range in 
higher orders. In such a manner, the exact GWF description is approached asymptotically
step by step. 

Here we apply the GWF solution to analyze the current approaches of strongly correlated electrons and single out the so-called $t$-$J$-$U$ model which may be regarded as an extended $t$-$J$ model with a relatively strong kinetic exchange and the direct Coulomb interactions included at the same time. Such combination of seemingly excluding each other processes requires a brief elaboration provided in the Appendix. However, one should note st the start, that as the exchange interactions are coming mainly from the interband $d$-$p$ processes, they are related only indirectly to the split-Hubbard-subband structure of 3$d$ states due to copper\cite{Zaanen1990,Eskes1993,Jefferson1992,Feiner1996,Avella2013,Emery1988,Valkov}.
As shown earlier, the $t$-$J$-$U$ model description leads to the antiferromagnetic (AF) phase stability for $\delta<1\%$\cite{Abram2013}, which is in rough agreement with experiment\cite{Keimer1992}. The appearance of both the AF exchange interaction and the direct Coulomb interaction within such approach brings into mind the competition between the spin-density-wave phase and the charge-density wave phase, the latter of which has been discovered in the cuprates recently\cite{Keimer2015,Tabis2014}. With the DE-GWF solution we not only reproduce the results of the variational quantum Monte Carlo calculations \cite{Randeria2012, Edegger2006}, sometimes with a better accuracy, but also carry out calculations for infinite systems within a reasonable computing time. This last factor allowed us to test a number of theoretical models ($t$-$J$, $t$-$J$-$U$, $t$-$J$-$U$-$V$, Hubbard) and single out the one which reproduces quantitatively the principal experimental data.

In brief, the principal aim of this paper is to confront the results obtained for the $t$-$J$-$U$ and related models with the experimental data for the HTS state in a proper quantitative manner. Explicitly, our purpose here is threefold: {\bf(i)} Not only to make a detailed comparison of selected experiments with theory, but first and foremost, to single out the universal characteristics such as Fermi velocity $v_F$, effective mass $m^{\star}$, Fermi wave vector $k_F$, and the non-BCS feature of the pairing, {\bf(ii)} Characterize whole class of theoretical single-band models based on strong correlations among the electrons and single out the one that allows for a quantitative predictions of selected dynamic properties (at least within the DE-GWF solution), and {\bf(iii)} To demonstrate the indispensability of the approach going beyond any current mean field approach (RMFT) on the example of DE-GWF.

From the formal point of view, we have reanalyzed the origin of the $t$-$J$ model\cite{Chao1976,Spalek1988,Anderson1988}. Namely, since the principal contribution to the antiferromagnetic exchange $J$ comes from superexchange via $2p$ states and the value of Hubbard interaction $U$ to the bare bandwidth is not too high, $U/W\sim 2.5-3$, we have extended the concept of $t$-$J$ model by incorporating expilictly the Coulomb interaction, allowing a small number of double occupancies, in addition to having a rather high value of $J$ what leads to the effective single-band model in the $t$-$J$-$U$ or even even $t$-$J$-$U$-$V$ form. The relevant microscopic interaction parameters of the starting model are schematically defined in Fig. \ref{fig:0}. The intersite Coulomb interaction $\sim V$ has been disregarded in the main text, but its role is elaborated briefly in the concluding Section. Here, we concentrate only on the quantitative analysis of the pure superconducting (SC) phase. Important issues which also can 
be tackled within 
the present approach, i.e., the description of electrodynamics in an applied magnetic field, are listed at the end.

\section{Model and method} 
The starting model is of the form of the extended Hubbard Hamiltonian with the antiferromagnetic exchange interaction\cite{Abram2013,Frantino2016,Zhang2003,SpalekOles}
\begin{equation}
\begin{split}
\mathcal{\hat{H}}&=\sideset{}{'}\sum_{ij\sigma}t_{ij}\hat{c}^{\dagger}_{i\sigma}\hat{c}_{j\sigma}+U\sum_i \hat{n}_{i\uparrow}\hat{n}_{i\downarrow}+ \sideset{}{'}\sum_{\langle ij\rangle}(V_{ij}-\frac{1}{4}J_{ij}) \hat{n}_{i}\hat{n}_{j}\\
&+ \sideset{}{'}\sum_{\langle ij\rangle}J_{ij}\hat{\mathbf{S}}_i\cdot\hat{\mathbf{S}}_j,
\end{split}
 \label{eq:H_start}
\end{equation}
where the primed summation means that $i\neq j$ and $\langle ij\rangle$ means that only pairs of nearest neighbors are taken into account. The first two terms represent the Hubbard model (consisting of the hopping and the intrasite repulsion terms, respectively), the third expresses the intersite Coulomb interaction (the part $\sim J_{ij}/4$ comes from the full expression for the exchange operator), and the last accounts for the antiferromagnetic exchange interaction (in the strict one-band representation of correlated electrons the exchange integral is $J_{ij}=2t^2_{ij}/(U-V_{ij})$ \cite{SpalekOles}). In the bulk of the paper we have disregarded the third term, as it does not influence much the quality of the comparison with the discussed here experimental data (see the discussion at the end of the paper). We should note that such a model has been introduced formally in Ref. \onlinecite{Zhang2003} as interpolating between the Hubbard- and $t$-$J$-model limits. The general form of the single-band model with 
all two-site interactions would require the inclusion of the pair-hopping and the so-called correlated hopping terms\cite{SpalekOles,Lauglin2014,Hirsh1989}. However, those two terms should be small as $U$ is relatively large; an additional check on their very small relevance, as well as of the three-site terms\cite{Spalek1988}, eventually comes from the quality of our quantitative description of selected experimental results.

The main task within our approach is to calculate the ground state energy and its properties for the full Gutzwiller-wave-function solution. This is carried out in a direct analogy to an earlier treatment of both the Hubbard\cite{Kaczmarczyk2013} and the t-J models\cite{Kaczmarczyk2014}. Explicitly, the ground state energy per lattice site is of the form 
\begin{equation}
 E_G\equiv\frac{\langle\Psi_G|\mathcal{\hat{H}}|\Psi_G\rangle}{N\langle\Psi_G|\Psi_G\rangle}\equiv\frac{1}{N}\langle\mathcal{\hat{H}}\rangle_G, 
 \label{eq:ground_state_energy}
\end{equation}
 where $N$ is the number of lattice sites, $|\Psi_G\rangle\equiv\hat{P}_G|\Psi_0\rangle$ is the Gutzwiller-type wave function, defined with the help of the operator $\hat{P}_G$ and the normalized uncorrelated state, $|\Psi_0\rangle$ (taken as the uncorrelated paired state with nonzero anomalous real space average $\langle\Psi_0|\hat{c}^{\dagger}_{i\uparrow} \hat{c}^{\dagger}_{j\downarrow}|\Psi_0 \rangle$, for $i\neq j$, when considering the SC phase). The $\hat{P}_G$  operator is of the form
 \begin{equation}
 \hat{P}=\prod_i\hat{P}_i=\prod_i\sum_{\Gamma}\lambda_{i,\Gamma} |\Gamma\rangle_{ii}\langle\Gamma|,
  \label{eq:Gutz_operator}
\end{equation}
where the variational parameters $\lambda_{i,\Gamma}\in\{\lambda_{i\emptyset},\lambda_{i\uparrow},\lambda_{i\downarrow},\lambda_{i d}\}$ correspond to four states from the local basis $|\emptyset\rangle_i\;, |\uparrow\rangle_i\;, |\downarrow\rangle_i\;, |\uparrow\downarrow\rangle_i$, respectively. In our analysis we assume spatial homogeneity, so $\lambda_{i,\Gamma}\equiv\lambda_{\Gamma}$. Moreover, we also limit to the spin-isotropic case, which means that $\lambda_{\uparrow}=\lambda_{\downarrow}=\lambda_1$.
 
 Within the diagrammatic expansion method\cite{Bunemann2012,Kaczmarczyk2013,Kaczmarczyk2014,Kaczmarczyk2015_2, Gebhard1990} one imposes the condition that 
 \begin{equation}
 \hat{P}_i^2\equiv1+x\hat{d}^{\textrm{HF}}_i,
 \label{eq:constraint}
 \end{equation}
 where $x$ is yet another variational parameter and $\hat{d}^{\textrm{HF}}_i=\hat{n}_{i\uparrow}^{\textrm{HF}}\hat{n}_{i\downarrow}^{\textrm{HF}}$, $\hat{n}_{i\sigma}^{\textrm{HF}}=\hat{n}_{i\sigma}-n_{0}$, with $n_{0}=\langle\Psi_0|\hat{n}_{i\sigma}|\Psi_0\rangle$. All the $\lambda_{\Gamma}$ parameters can be expressed with the use of the $x$ parameter due to (\ref{eq:constraint}) and (\ref{eq:Gutz_operator}), which means that we are left with only one variational parameter in the considered case.
  
 The expectation values of the consecutive terms which appear in the $t$-$J$-$U$ Hamiltonian (\ref{eq:H_start}) (we omit the intersite Coulomb repulsion term here) can be expressed in the form of the power series
 \begin{equation}
\begin{split}
  \langle\Psi_G|\hat{c}^{\dagger}_{i\sigma}\hat{c}_{j\sigma}|\Psi_G\rangle&=\sum_{k=0}^{\infty}\frac{x^k}{k!}\sideset{}{'}\sum_{l_1...l_k}\langle \tilde{c}^{\dagger}_{i\sigma}\tilde{c}_{j\sigma}\hat{d}^{\textrm{HF}}_{l_1...l_k} \rangle_0,\\
  \langle\Psi_G|\hat{s}^{\dagger}_{i\sigma}\hat{s}_{j\bar{\sigma}}|\Psi_G\rangle&=\lambda_{1}^4\sum_{k=0}^{\infty}\frac{x^k}{k!}\sideset{}{'}\sum_{l_1...l_k}\langle \hat{s}^{\dagger}_{i\sigma}\hat{s}_{j\bar{\sigma}}\hat{d}^{\textrm{HF}}_{l_1...l_k} \rangle_0,\\
  \langle\Psi_G|\hat{d}_i|\Psi_G\rangle&=\lambda_d^2\sum_{k=0}^{\infty}\frac{x^k}{k!}\sideset{}{'}\sum_{l_1...l_k}\langle \hat{d}_i\hat{d}^{\textrm{HF}}_{l_1...l_k} \rangle_0,\\
\end{split}
\label{eq:expectation_val_terms}
\end{equation}
where $\hat{s}_{i\sigma}=\hat{c}^{\dagger}_{i\sigma}\hat{c}_{i\bar{\sigma}}$, $\tilde{c}^{(\dagger)}_{i\sigma}=\hat{P}_i\hat{c}^{(\dagger)}_{i\sigma}\hat{P}_i$, and  $\hat{d}^{\textrm{HF}}_{\varnothing}\equiv 0$. The primmed summation on the right hand side has the restrictions $l_p\neq l_{p'}$, $l_p\neq i,j$ for all $p$, $p'$. Next, by using the Wicks theorem the non-correlated averages in Eqs. (\ref{eq:expectation_val_terms}) can be expressed in terms of $P_{ij}\equiv\langle \Psi_0|\hat{c}^{\dagger}_{i\sigma} \hat{c}_{j\sigma}|\Psi_0\rangle$ and $S_{ij}\equiv\langle \Psi_0| \hat{c}^{\dagger}_{i\uparrow} \hat{c}^{\dagger}_{j\downarrow}| \Psi_0\rangle$. Due to the fact that the Gutzwiller operator may change the norm of the noncorrelated wave function, one has to divide the above expressions by $\langle\Psi_G|\Psi_G\rangle$, while calculating the ground state energy. 
It is convenient to use the linked-cluster theorem\cite{Kaczmarczyk2014, Fetter} to simplify the expressions obtained in the described manner. Such approach allows us to evaluate the ground state energy to a sufficient accuracy by including the first $4-6$ orders of the diagrammatic expansion\cite{Bunemann2012}, depending on the model at hand. 
 
From the minimization condition of the ground state energy (\ref{eq:ground_state_energy}) one can derive the effective Hamiltonian, which for the case of pure superconducting phase has the form
%It is convenient to derive the effective Hamiltonian representing the results which correspond to the minimum of the ground state energy (\ref{eq:ground_state_energy}). For the case of pure superconducting phase such Hamiltonian has the following form
\begin{equation}
 \hat{\mathcal{H}}_{\textrm{eff}}=\sum_{ij\sigma}t^{\textrm{eff}}_{ij}\hat{c}^{\dagger}_{i\sigma}\hat{c}_{j\sigma}+\sideset{}{'}\sum_{ij}\big(\Delta^{\textrm{eff}}_{ij}\hat{c}^{\dagger}_{i\uparrow}\hat{c}^{\dagger}_{j\downarrow}+H.c.\big),
 \label{eq:H_effective}
\end{equation}
where the effective hopping and the effective superconducting gap parameters are defined through the corresponding relations
\begin{equation}
 t^{\textrm{eff}}_{ij}\equiv \frac{\partial\mathcal{F}}{\partial P_{ij}},\quad \Delta^{\textrm{eff}}_{ij}\equiv \frac{\partial\mathcal{F}}{\partial S_{ij}}.
 \label{eq:effective_param}
\end{equation}
For $i=j$, the $t^{\textrm{eff}}_{ij}$ has an interpretation of an effective chemical potential. The expression for ground state energy functional per atomic site $\mathcal{F}=E_G-\mu_Gn_G$ (where $\mu_G$ and $n_G$ are the chemical potential and the number of particles per lattice site determined in the correlated state, respectively) is obtained via the corresponding diagrammatic expansions of all the averages contained in it. It should be noted that within this approach one can also calculate the correlated superconducting gap parameter $\Delta_G=\langle \hat{c}^{\dagger}_{i\uparrow} \hat{c}^{\dagger}_{j\downarrow}\rangle_G$, which is analyzed in the main text. Namely, the correlated and the effective gaps are expressed through the uncorrelated quantities $P_{ij}$, $S_{ij}$, and the variational parameter $x$. In result, one has to solve a set of integral equations for $P_{ij}$,
 $S_{ij}
$, $\mu_G$, and $x$, from which the electronic structure, the 
correlated gap $\Delta_G$, and the ground state energy are explicitly evaluated for given set of parameters $t_{ij}$, $U$, $J$, and the band filling $n_G\equiv(1-\delta)$.

Some methodological remarks are in place here. Namely, the application of the Wicks theorem allows us to express the elements of the sums in (\ref{eq:expectation_val_terms}) as diagrams with the lattice sites playing the role of vertices and $P_{ij}$, $S_{ij}$ being the edges connecting those vertices. In the obtained diagrammatic sums the site indices $l_1...\;l_k$ run over all the lattice sites. However, the $P_{ij}$ and $S_{ij}$ with significantly large distance $|\Delta \mathbf{R}|=|\mathbf{R}_i-\mathbf{R}_j|$ lead to small contributions\cite{Kaczmarczyk2015_2}. Therefore, during the calculations one may limit to terms with lines that correspond to distances smaller than some $R_{\textrm{max}}$. Moreover, it is convenient to introduce additional condition that a given diagram contribution is included in the calculations if the sum of all the lines length (in the Manhattan metrics) which correspond 
to this diagram is smaller than some 
specified 
value $R_s$. 

Furthermore, beginning from some particular value of the expansion order $k=k_{\textrm{max}}$ one can neglect the terms of the summation in Eqs. (\ref{eq:expectation_val_terms}). In such situation one includes diagrams with number of vertices up to $k_{\textrm{max}}+1$ ($k_{\textrm{max}}+2$) corresponding to one- (two-) site terms of the Hamiltonian (\ref{eq:H_start}). The order of the approach is then equal to $k_{\textrm{max}}$. In practice, it is convenient to include diagrams with number of lines up to some particular value, $N_l$. It should be noted that including only the zeroth-order diagrams leads to calculations which are equivalent to the SGA version of the renormalized mean-field theory.

All the presented results have been obtained for the calculation parameters set to $R^2_{\textrm{max}}=10$, $R_l=26$, and $N_l=13$. The value of $N_l=13$ means that we include diagrams up to the fifth order and some additional diagrams which are of the sixth order. The set of integral equations for $P_{ij}$, $S_{ij}$, $\mu_G$, and $x$ has been solved with the use of GSL numerical library with the typical accuracy set to $10^{-7}$.

\section{Results}

In our analysis we take into account both the nearest-neighbor and the next-nearest-neighbor hoppings, with the respective hopping integrals set to $t=-0.35$ eV (the value $|t|$ is taken as the energy unit, if not specified explicitly) and $t'=0.25|t|$. The intersite exchange integral is assumed as nonzero only for the nearest-neighbors $\langle i,j \rangle$, $J_{<ij>}\equiv J$. Below we analyze concrete measurable quantities and compare them o experiment in a quantitative manner.
%%%%%%%%%%%%%%%%%%%%%%%%%%%%%%%%%%%%%%%%%%%%FIG0%%%%%%%%%%%%%%%%%%%%%%%%%%%%%%%%
\begin{figure}[h!]
%\hfill
\centering 
\epsfxsize=75mm 
%{\epsfbox[29 114 1495 1319]{fig000_m.eps}}
{\epsfbox[29 114 982 846]{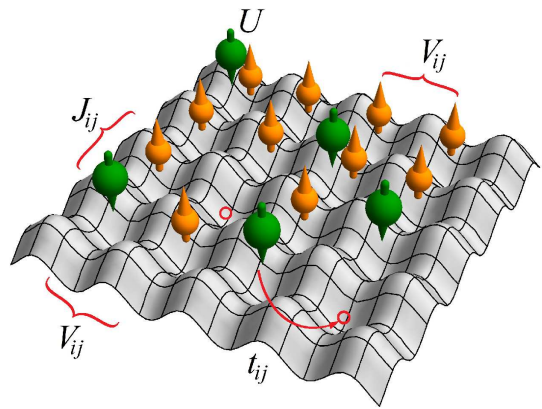}}
\caption{Schematic representation of the single plane of Cu ions with active electrons located on the 3$d_{x^2-y^2}$ orbitals and 'dressed' with 2$p_{\sigma}$ orbitals (not shown). The microscopic parameters of such single narrow-band model of correlated electrons are: (i) hopping integral $t_{ij}=t=-0.35$ eV for nearest neighbors hopping and $t_{ij}=t'=0.25|t|$ for next-nearest neighbors hopping, (ii) intraatomic (Hubbard) Coulomb interaction $U\approx8$ eV, (iii) interatomic (intersite) Coulomb interaction $V_{ij}$ (omitted in the main text of the paper), and (iv) antiferromagnetic exchange integral $J_{ij}\sim70-80$ meV. The empty sites (with no electrons, red circles) are called holes, with $\delta$ characterizing the average probability of their appearance per site.}
\label{fig:0}
\end{figure}
%%%%%%%%%%%%%%%%%%%%%%%%%%%%%%%%%%%%%%%%%%%%%%%%%%%%%%%%%%%%%%%%%%%%%%%%%%%%%%%%
\subsection{Kinetic energy gain and condensation energy}
\begin{figure}[!ht]
%\hfill
\centering
\epsfxsize=70mm 
%{\epsfbox[55 382 561 770]{fig2.eps}}
{\epsfbox[20 18 485 405]{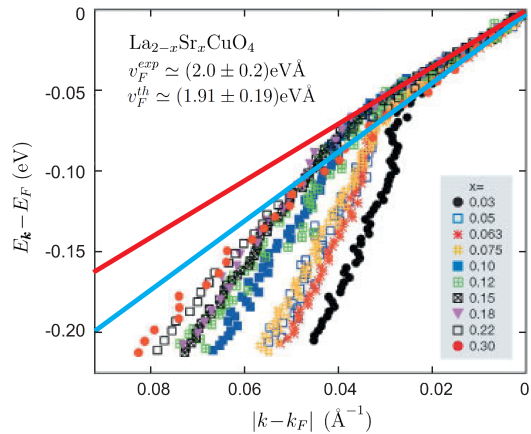}}
\caption{Universal Fermi velocity $v_F$ in the nodal direction for La$_{2-x}$Sr$_x$CuO$_4$ as determined from the linear part of the dispersion relation $E_{\mathbf{k}}$ (data taken from Ref. \onlinecite{Zhou2003}). The slopes of the straight lines determine $v_F$ in the extreme cases. The extracted from the data mean value $v^{exp}_F=(2.0\pm0.2)$ eV\AA, $\;$ is in very good accord with our theoretical result $v^{th}_F=(1.91\pm0.19)$ eV\AA$\;$ obtained within the full GWF (see main text). The stoichiometry parameter $x$ in the legend characterizes the hole concentration $\delta$.}
\label{fig:1}
\end{figure}

To set the stage-reference point of our analysis in Fig. \ref{fig:1} we show the dispersion relation obtained experimentally for La$_{2-x}$Sr$_x$CuO$_4$ close to the Fermi energy according to Refs.\cite{Zhou2003,Shrakopi2008}. The universal Fermi velocity $v_F$ in the nodal direction is estimated from the data by taking the slopes of the extreme curves as marked by the straight lines, what leads to the average result $v_F^{exp}=(2.0\pm0.2)$ eV\AA $\;$(in these units $v_F \equiv \hbar v_F$, in physical units $v_F\approx 2\cdot 10^{7}$ cm/s). This value is in very good agreement with the one determined theoretically here (see below) $v_F^{th}=(1.91\pm0.19)$ eV\AA, what illustrates the quality of our approach, the results of which we discuss in detail next.
%%%%%%%%%%%%%%%%%%%%%%%%%%%%%%%%%%%%%%%%%%%%FIG1%%%%%%%%%%%%%%%%%%%%%%%%%%%%%%%%
%%%%%%%%%%%%%%%%%%%%%%%%%%%%%%%%%%%%%%%%%%%%%%%%%%%%%%%%%%%%%%%%%%%%%%%%%%%%%%%%

To single out the model which describes properly the high temperature superconductivity, we have analyzed the Hubbard, $t$-$J$, $t$-$J$-$U$, and $t$-$J$-$U$-$V$ models separately, all within the full Gutzwiller wave function solution (the parameters are visualized in Fig. \ref{fig:0}). As we show below, the $t$-$J$-$U$ model reproduces the universal characteristics quantitatively. 
%%%%%%%%%%%%%%%%%%%%%%%%%%%%%%%%%%%%%%%%%%%%FIG2%%%%%%%%%%%%%%%%%%%%%%%%%%%%%%%%
\begin{figure}[!ht]
%\hfill
\centering
\epsfxsize=65mm 
%{\epsfbox[0 0 420 316]{fig3.eps}}
%{\epsfbox[10 10 421 315]{fig3.eps}}
{\epsfbox[194 382 429 678]{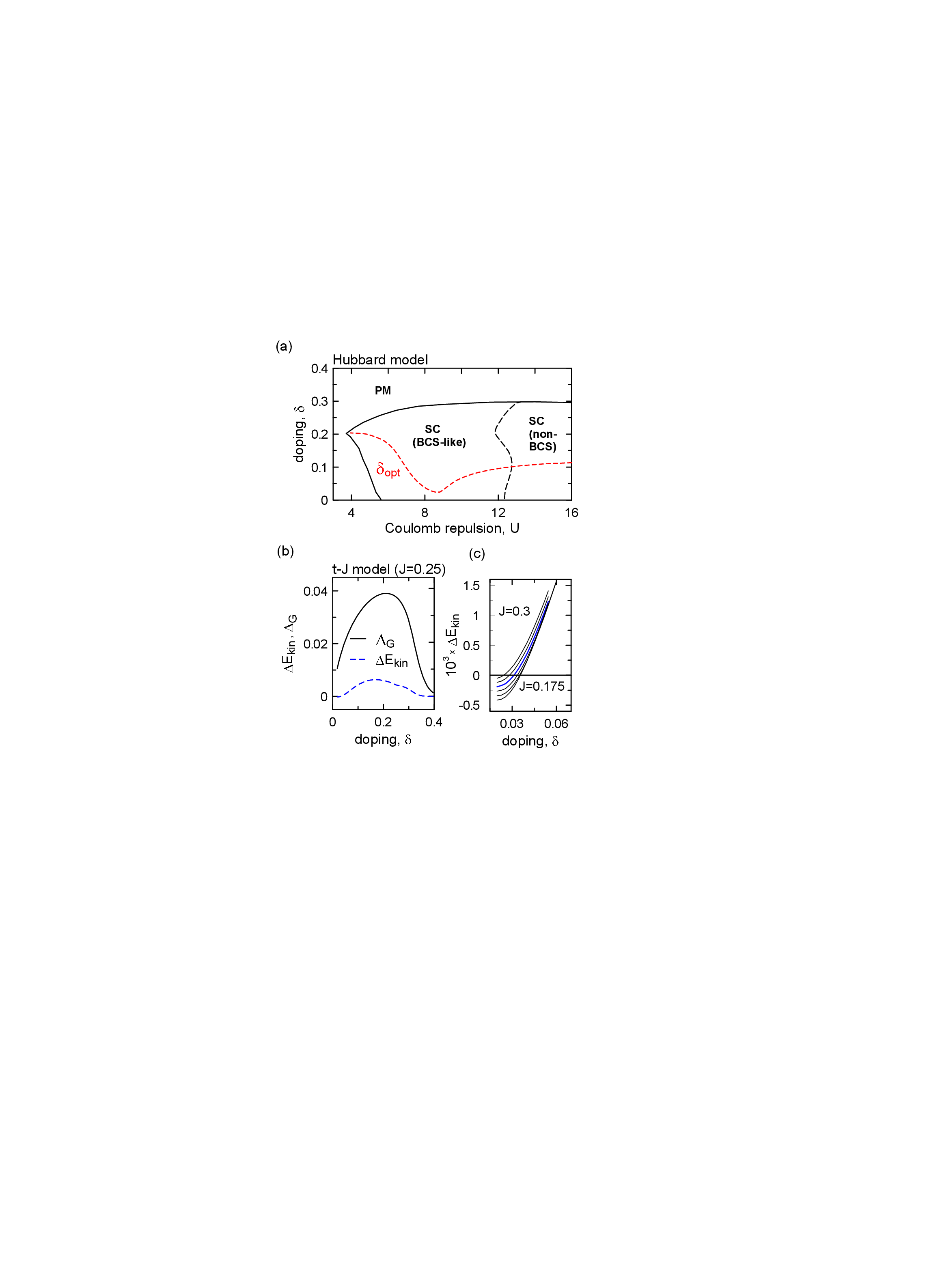}}
\caption{(a) The border between the BCS-like and non-BCS superconducting regimes (with
$\Delta E_{kin}>0$ and $\Delta E_{kin}<0$, respectively) for the Hubbard model (the red dashed line provides the optimal doping, $\delta_{\textrm{opt}}$ vs $U$). (b) Correlated gap ($\Delta_G$) and the kinetic energy loss $\Delta E_{kin}$, both vs doping for the t-J model with $J=0.25$. In (c) we show that the kinetic energy gain ($\Delta E_{kin}<0$) appears also for the case of the $t$-$J$ model but very close to half-filling and the non-BCS region slightly broadens with decreasing $J$. The subsequent curves represent changing the $J$ value by $0.025$ between $J=0.3$ and $J=0.175$ with the blue one corresponding to $J=0.25$.}
\label{fig:2} 
\end{figure}
%%%%%%%%%%%%%%%%%%%%%%%%%%%%%%%%%%%%%%%%%%%%%%%%%%%%%%%%%%%%%%%%%%%%%%%%%%%%%%%%

%%%%%%%%%%%%%%%%%%%%%%%%%%%%%%%%%%%%%%%%%%%%FIG2%%%%%%%%%%%%%%%%%%%%%%%%%%%%%%%%
\begin{figure}[!ht]
%\hfill
\centering
\epsfxsize=75mm 
%{\epsfbox[0 0 420 316]{fig3.eps}}
%{\epsfbox[10 10 421 315]{fig3.eps}}
{\epsfbox[10 10 158 176]{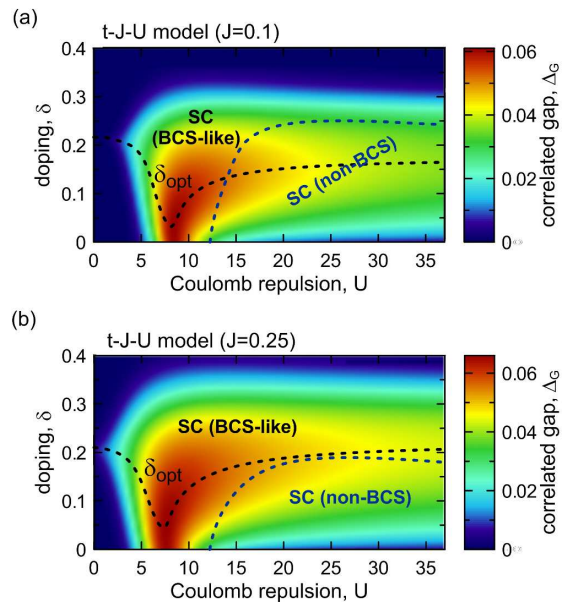}}
\caption{(a) and (b), BCS-like and non-BCS superconducting states for the t-J-U model for $J=0.1$ and $J=0.25$, respectively. The latter value matches the experiment (see main text). Note that in (b) the non-BCS state appears only in the underdoped regime.}
\label{fig:2_1} 
\end{figure}
%%%%%%%%%%%%%%%%%%%%%%%%%%%%%%%%%%%%%%%%%%%%%%%%%%%%%%%%%%%%%%%%%%%%%%%%%%%%%%%%

%The kinetic energy gain has been considered before within the cluster dynamic mean-field theory (CDMFT),
%where it has been shown to change sign at doping $\delta\backsimeq 0.15$. To fit the experimental data quantitatively in Fig. \ref{fig:2} a-e...

In Figs. \ref{fig:2} a-c and \ref{fig:2_1} a,b we discuss one of the principal non-BCS features of the SC state, namely the kinetic energy gain in the superconducting state with respect to the normal paramagnetic (PM) state\cite{Deutscher2005,Carbonne2006,Molegraaf2002,Gianetti2011}. This gain is defined as
\begin{equation}
 \Delta{E}_{\textrm{kin}}\equiv E^{SC}_{G|0}-E^{PM}_{G|0},\quad E_{G|0}\equiv \frac{1}{N}\sideset{}{'}\sum_{ij\sigma}t_{ij}\langle\hat{c}^{\dagger}_{i\sigma}\hat{c}_{j\sigma}\rangle_G,
\end{equation} 
where the kinetic energy difference is taken between the SC and PM states. Subscript 'G' means that the average $\langle...\rangle$ is taken in the Gutzwiller state $|\Psi_G\rangle$. Note that the condensation energy, corresponding to the total ground-state-energy difference, $\Delta E_c\equiv E^{SC}_G-E^{PM}_G$, is always negative for the SC phase to be stable. In Fig. \ref{fig:2}a we display the results for the Hubbard model concerning the stability of the $d$-wave SC with respect to the normal (PM) state on the plane doping $\delta$ - intraatomic Coulomb repulsion $U$. The BCS-like (with $\Delta E_{\mathrm{kin}}>0$) and non-BCS (with $\Delta 
E_{\mathrm{
kin}}<0$) 
regimes  
are 
separated in this 
case by an almost vertical dashed line which 
illustrates the fact that the latter regime appears as stable if only $U\gtrsim 12$. The optimal doping (i.e.,
 the 
doping with the maximal value of the transition temperature $T_c$) is denoted by $\delta_{
\textrm{opt}}$ and for a given $U$ is determined theoretically by taking the value of the doping which corresponds to maximal correlated gap, $\Delta_G\equiv \langle \hat{c}^{\dagger}_{i\uparrow}\hat{c}^{\dagger}_{j\downarrow}\rangle_G$ for $\mathbf{R}_i-\mathbf{R}_j=(1,0)a$ (where $a$ is the lattice constant). For comparison, in Fig. \ref{fig:2} b and c the results for $\Delta E_{\mathrm{kin}}$ and the magnitude of the gap are displayed vs $\delta$ for the case of $t$-$J$ model. As shown in Fig. \ref{fig:2}c, the non-BCS ($\Delta E_{\mathrm{kin}}$<0) state appears only close to half filling for this model. However, it should be noted that the result differs from those obtained within the cluster DMFT \cite{Haule2007}. As one can see, the decreasing of the value of $J$ leads to the appearance of the non-BCS behavior for slightly larger dopings, $\delta$. Nonetheless, it is impossible to fit the experimental data\cite{Deutscher2005} with a reasonable value of $J$. One should note that for $J=0$ and $d^2=0$ 
the 
non-BCS behaviour appears in 
the whole doping range of the paired state stability, as we show at the end of this Section (cf. Fig. 
\ref{fig:4}b). It would be important to carry out a detailed comparison between the CDMFT \cite{Carbonne2006} results and those presented here. Note that none of the results shown in Figs. \ref{fig:2}a-c do reflect the 
proper behavior of the experimental data, according to which the non-BCS behavior appears up to almost optimal doping\cite{Deutscher2005,Gianetti2011,Molegraaf2002,Carbonne2006} (as we also show explicitly below). In Fig. \ref{fig:2_1}a and b we exhibit the phase diagram for the $t$-$J$-$U$ model for two values of the exchange integral, $J=0.1$ and $J=0.25$, respectively. As can be seen, by including both $U$ and $J$ simultaneously one obtains the 
transition from the non-BCS to the BCS-like regime very close to the optimal doping ($\delta_
{
\mathrm{opt}}$) for proper values of the model parameters. Moreover, for $J=0.25$ the non-BCS region appears only up to the optimally doped case.
%%%%%%%%%%%%%%%%%%%%%%%%%%%%%%%%%%%%%%%%%%%FIG3%%%%%%%%%%%%%%%%%%%%%%%%%%%%%%%%
\begin{figure}[!ht]
%\hfill 
\centering
\epsfxsize=70mm 
{\epsfbox[232 475 485 777]{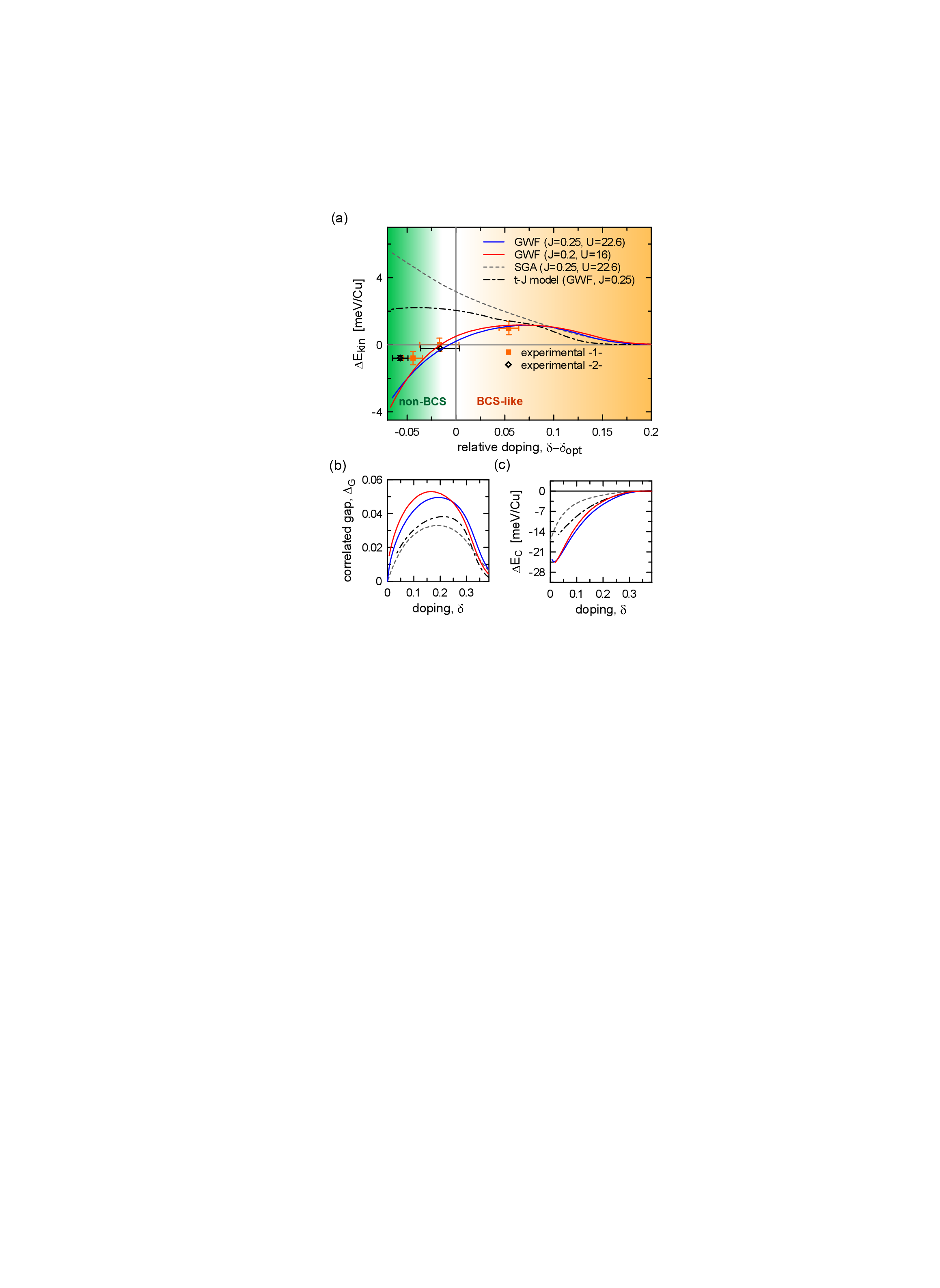}}
\caption{(a) Kinetic energy gain $\Delta E_{\mathrm{kin}}$ vs relative doping $\delta-\delta_{\mathrm{opt}}$ ($\delta_{\mathrm{opt}}$ is the optimal doping). Theoretical curves have been obtained for the $t$-$J$-$U$ model within the GWF solution for $J=0.25$, $U=22.6$ (blue solid line) and for $J=0.2$, $U=16$ (red solid line), while the experimental points are taken from Ref. \onlinecite{Deutscher2005}. For comparison, the results for the same model within the SGA method (gray dashed line)  and for the $t$-$J$ ($J=0.25$) model in GWF methodology (dot-dashed curve) are also drawn. Note that only the GWF solution of the $t$-$J$-$U$ model reproduces quantitatively the experimental data. (b) and (c) Correlated gap magnitude $\Delta_G$ and the condensation energy $\Delta E_C$, both as a function of doping, are drawn for the same values of the parameters and the same models.}
\label{fig:3}
\end{figure}  
%%%%%%%%%%%%%%%%%%%%%%%%%%%%%%%%%%%%%%%%%%%%%%%%%%%%%%%%%%%%%%%%%%%%%%%%%%%%%%%%

%%%%%%%%%%%%%%%%%%%%%%%%%%%%%%%%%%%%%%%%%%%FIG3%%%%%%%%%%%%%%%%%%%%%%%%%%%%%%%%
\begin{figure}[!ht]
%\hfill
\centering
\epsfxsize=75mm 
{\epsfbox[207 462 461 650]{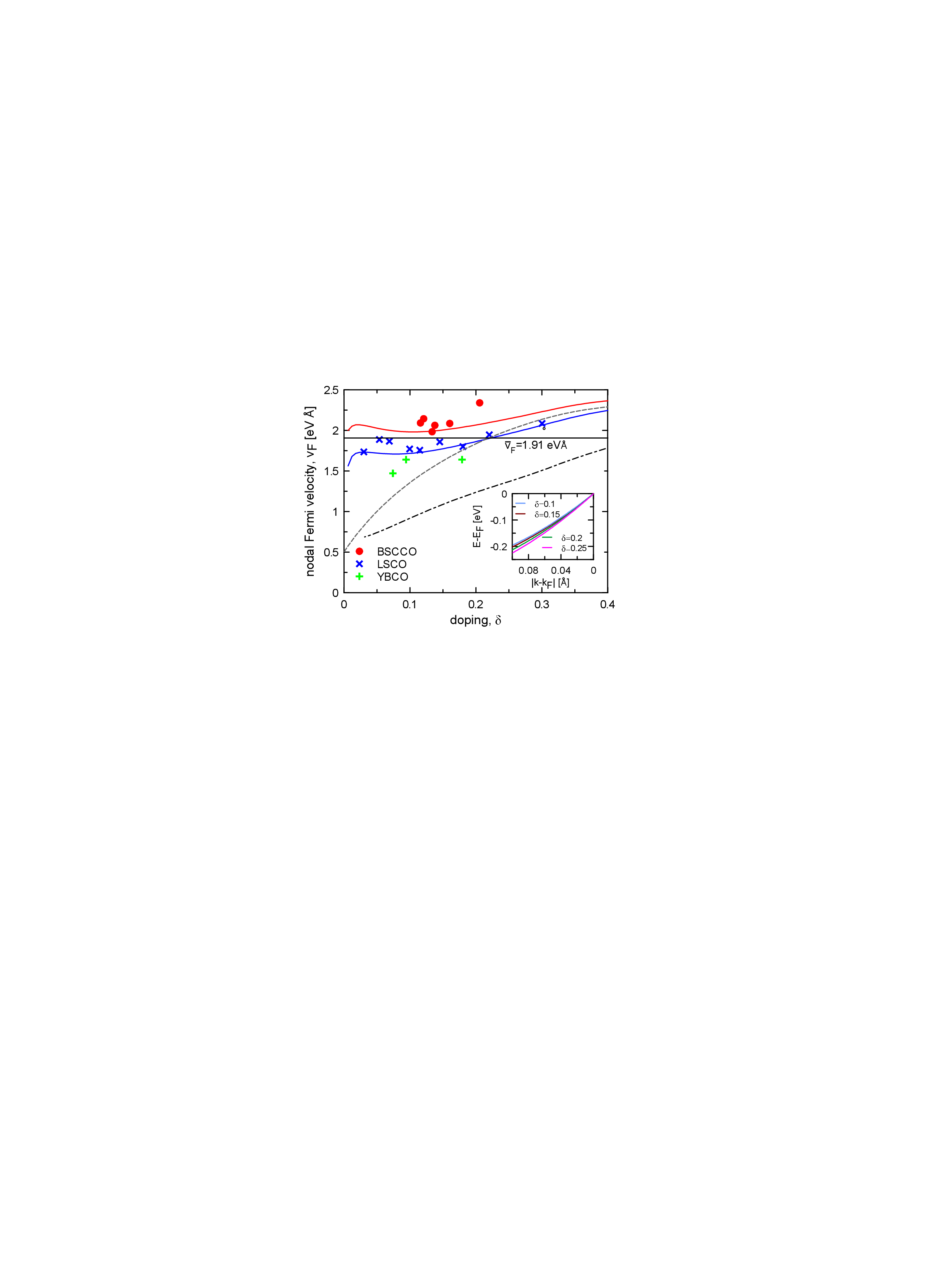}}
\caption{Fermi velocity in the nodal direction ($k_x=k_y$) versus $\delta$; the theoretical curves are drawn for the 
same parameter values and labeling as that in Fig. \ref{fig:3}a. The experimental data are taken from Ref. \cite{
Kordyuk2005} (BSCCO), Ref. \cite{Zhou2003} (LSCO), and Ref. \onlinecite{Borisenko2006} (YBCO). Note that only the GWF solution of the $t$-$J$-$U$ model reproduces quantitatively the experimental data . The representive theoretical value $v_F=(1.91\pm0.19)$ eV\AA$\;$ listed in Fig. \ref{fig:1} was obtained by fitting a horizontal line $v_F=const$ to our results for LSCO, as marked. In the inset we show the theoretical dispersion relations close to $\mathbf{k}_F$ for selected $\delta$-values. The calculated curves do not contain any abrupt change in the dispersion relations at $\sim 80$ meV seen in the experiment (cf. Fig. \ref{fig:1}) and ascribed to a strong electron-lattice coupling \cite{Lanzara2001}.}
\label{fig:3_1}
\end{figure}  
%%%%%%%%%%%%%%%%%%%%%%%%%%%%%%%%%%%%%%%%%%%%%%%%%%%%%%%%%%%%%%%%%%%%%%%%%%%%%%%%

%%%%%%%%%%%%%%%%%%%%%%%%%%%%%%%%%%%%%%%%%%%FIG3%%%%%%%%%%%%%%%%%%%%%%%%%%%%%%%%
\begin{figure}[!ht]
%\hfill
\centering
\epsfxsize=70mm 
{\epsfbox[101 494 301 745]{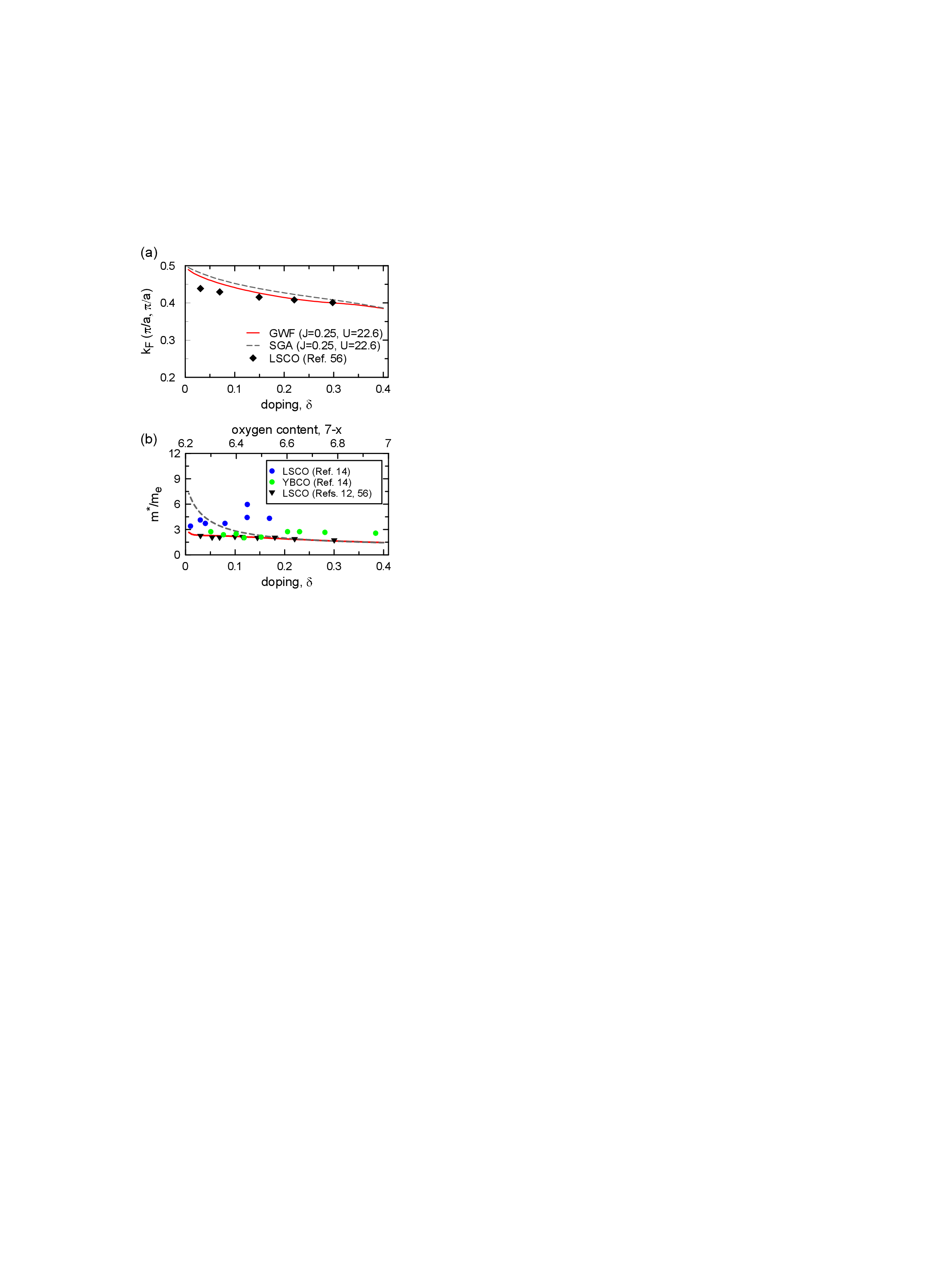}}
\caption{(a) Fermi momentum $k_F$ as a function of doping for the case of $t$-$J$-$U$ model within the GWF solution (red solid line) and for the SGA method (dashed line), compared with the experimental data for LSCO taken from Ref. \onlinecite{Hashimoto2008}. (b) Relative electron effective mass in the nodal direction as a function of doping calculated for the same approaches and parameters as in (a), compared with the experimental data for LSCO and YBCO\cite{Padilla2005}. We also present the effective mass values obtained by using the measured Fermi velocity for LSCO (taken from Ref. \onlinecite{Zhou2003}) and the Fermi momentum (taken from Ref. \onlinecite{Hashimoto2008}). The experimental data for YBCO (green dots) are presented as a function of the oxygen content (top axis) while the data for LSCO (blue dots and black inverted triangles) and the theoretical lines are presented as a function of hole concentration (bottom axis).}
\label{fig:35}
\end{figure}  
%%%%%%%%%%%%%%%%%%%%%%%%%%%%%%%%%%%%%%%%%%%%%%%%%%%%%%%%%%%%%%%%%%%%%%%%%%%%%%%%

The actual behavior of the data concerning the non-BCS regime appearance\cite{Deutscher2005} is displayed in Fig. \ref{fig:3}a, where the blue and red continuous lines represent the GWF solution for $t$-$J$-$U$ model for ($J=0.25$, $U=22.6$) and ($J=0.2$, $U=16$), respectively, whereas the experimental data are taken from Ref.\cite{Deutscher2005}. For the sake of comparison, we plot the corresponding results (the full GWF solution) obtained for the $t$-$J$ model (dot-dashed line) and those for the $t$-$J$-$U$ model within the SGA method (dashed line), which is a more sophisticated form of the RMFT. As one can see, we have obtained a good agreement with the experiment for the case of the $t$-$J$-$U$ model only for the solution going beyond the RMFT method. In Fig. \ref{fig:3}b and c we present, respectively, the values of the correlated real-space gap $\Delta_G$ and the 
condensation energy $\Delta E_c$ for 
all the approaches considered in Fig. \ref{fig:3}a and with the same values of respective parameters characterized by the corresponding colors of the curves. The superconducting state with the gap magnitude $\Delta_G>10^{-4}$ ($>0.5K$) persists up 
to the doping $\delta_{c2}\simeq 0.43$, which is still substantially 
larger than the observed value $\sim 0.3$. The inclusion of a weak intersite Coulomb interaction ($V\lesssim 1$) diminishes $\delta_{c2}$ to the experimental value. However, the influence of the last factor must be discussed in conjunction with a detailed analysis of other phases \cite{Kordyuk2005} as such investigation involves a delicate balance of multiple coexisting orderings (ferromagnetism, charge-density wave). 

\subsection{Universal Fermi velocity}
In Fig. \ref{fig:3_1} we plot the doping dependence of the Fermi velocity in the nodal direction for the case of the $t$-$J$-$U$ model (blue and red solid lines), as well as the results coming from either the full solution of the $t$-$J$ model (dot-dashed line) and from the renormalized mean-field (SGA) solution of the $t$-$J$-$U$ model (dashed line). Again, only the full GWF solution of the $t$-$J$-$U$ model represents quantitatively the data trend for La$_{2-\delta}$Sr$_{\delta}$CuO$_4$ (LSCO)\cite{Zhou2003} and Bi$_2$Sr$_2$CaCu$_2$O$_8$ (BSCCO)\cite{Kordyuk2005}, though the results of GWF and SGA coalesce in the overdoped region, which can be regarded as a universal BCS-like limit. For completeness, we have added few known points for the YBCO\cite{Borisenko2006}, as marked by green crosses in the Figure. The data for YBCO and BSCCO are scarce, but the values are still close to those for LSCO illustrating the universality of the $v_F$ value. Furthermore, we fit a line $v_F=const$ to our results for LSCO, 
to 
obtain the overall value $v^{th}_F=(1.91\pm 0.19)$ eV\AA, provided already in Fig. \ref{fig:1} which agrees also very good with the value $v^{exp}_F=(2.0\pm 0.2)$eV\AA$\;$ obtained from independent experiments\cite{Zhou2003,Shrakopi2008}. These two sets of data are not only consistent 
with each other but also provide a strong support for the interpretation of the $v_F$ universality. Note that the fits in Figs. \ref{fig:3}a and \ref{fig:3_1} for LSCO have been carried out for the same set of the parameters: $t=-0.35$eV, $t'=0.25|t|$, $J=0.25|t|$, $U=22.6|t|$ (blue solid lines).
%%%%%%%%%%%%%%%%%%%%%%%%%%%%%%%%%%%%%%%%%%%%FIG4%%%%%%%%%%%%%%%%%%%%%%%%%%%%%%%%
\begin{figure}[!ht]
%\hfill 
\centering 
\epsfxsize=75mm 
{\epsfbox[206 442 450 741]{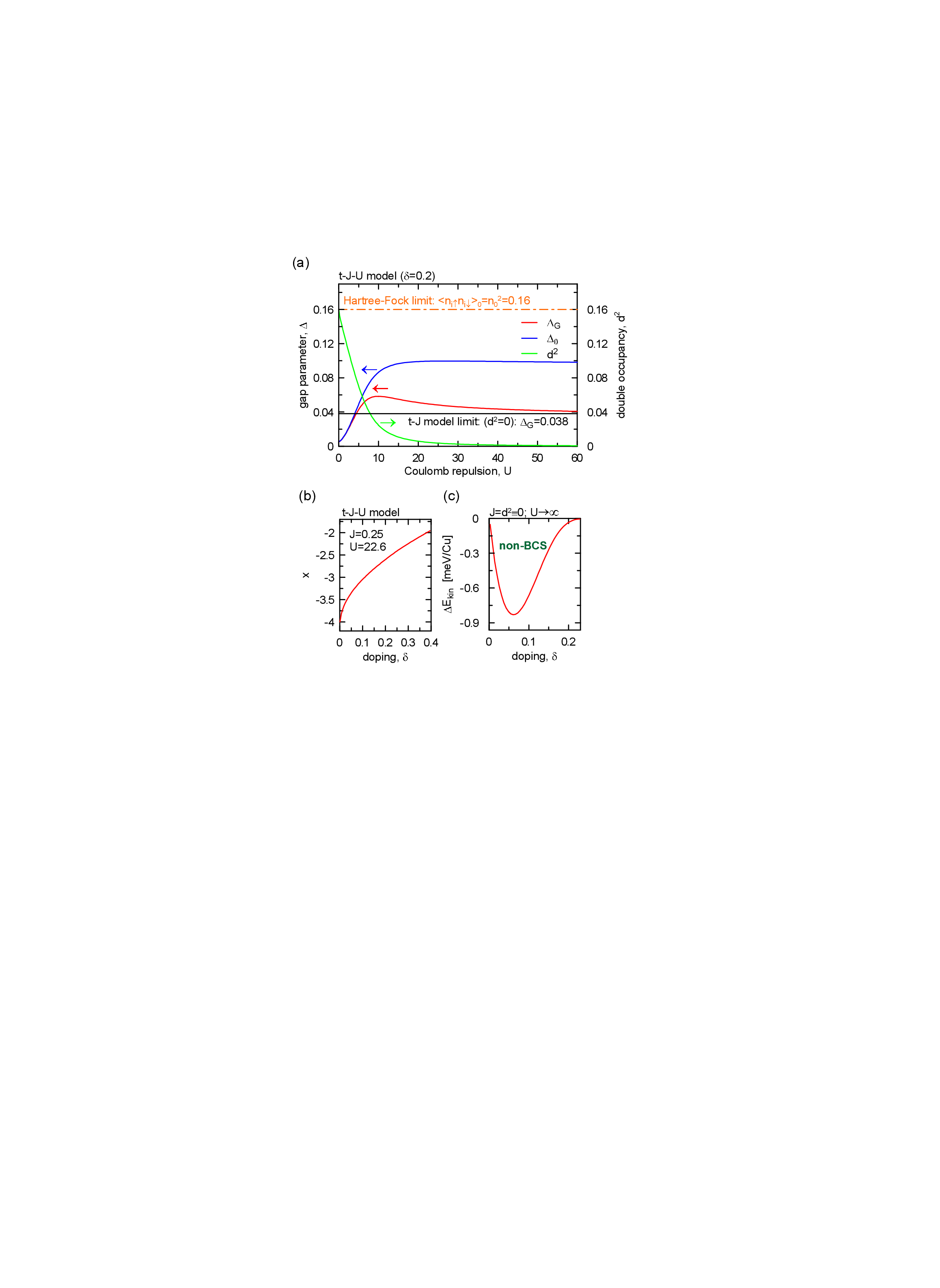}}
\caption{(a), Correlated ($\Delta_G$) and the noncorrelated ($\Delta_0$) gaps, as well as the double occupancy probabilities in the correlated ($d^2$) and uncorrelated ($n^2_0$) states, all as functions of the Hubbard interaction magnitude $U$. Note that with the increasing $U$ the double occupancy tends to $d^2=0$ and the correlated gap approaches the value $\Delta_G=0.038|t|$, which corresponds to that obtained in the $t$-$J$ model (i.e., when $d^2\equiv0$). Also, when the Coulomb interaction is weak, the correlated and uncorrelated values of the SC gap coalesce as it should be. (b) The variational parameter $x$ (cf. Eq.(\ref{eq:constraint})) as a function of doping for the case of the $t$-$J$-$U$ model with $J=0.25$ and $U=22.6$. Note that as the doping approaches half-filling ($\delta\rightarrow 0$) the $x$ parameter tends to the value $-4$ which corresponds to the $t$-$J$ model case with no double occupancies ($d^2=0\rightarrow x=-1/(1-n_0)^2$, cf. Ref. \onlinecite{Kaczmarczyk2014}).
(c) The kinetic energy gain vs $\delta$ for the Hubbard model with $U\rightarrow \infty$ and $J=0$. The curve is similar to that in Fig. \ref{fig:3}c since here $\Delta E_c\equiv \Delta E_{\mathrm{kin}}$, but the actual values are now an order of magnitude smaller showing that the kinetic energy gain alone in the $U\rightarrow 0$ limit cannot 
explain the observed $T_c$ magnitude for HTS.}
\label{fig:4}
\end{figure}
%%%%%%%%%%%%%%%%%%%%%%%%%%%%%%%%%%%%%%%%%%%%%%%%%%%%%%%%%%%%%%%%%%%%%%%%%%%%%%%%
From the above results concerning $v_F$ independence of $\delta$ one can draw a very important conclusion. Namely, the electronic structure in the nodal direction has a universal character in the sense that it survives the effect of the shrinking of the Fermi surface with the diminishing carrier (hole) concentration. Note also that the value of $v_F$ is the same, independently of the circumstance that other phases or pseudogap may appear in the system. Thus, these subsidiary phenomena must also have a gap node in that direction, in which $v_F$ has been determined, so they do not influence directly the dispersion-relation gradient at $E_F$.

\subsection{Fermi wave vector and effective mass}
In Fig. \ref{fig:35}a we provide a direct comparison of the experimentally determined\cite{Hashimoto2008} value of the Fermi wave vector $k_F\equiv(k_x^2+k_y^2)^{1/2}$ (in units of $\sqrt{2}\pi/a$), with the theoretical results: GWF (red solid line) and SGA (dashed line). Both of the approaches provide a correct trend, with a slight systematic deviation for $\delta\lesssim 0.1$ with the GWF results being closer to the experiment. In Fig. \ref{fig:35}b we present the calculated effective mass enhancement in the nodal direction by using the full GWF solution and the SGA approximation (red solid line and gray dashed line, respectively) both as a function of doping in comparison with the corresponding values for LSCO (blue dots) and YBCO (green dots) measured by a combination of dc transport and infrared spectroscopy (taken from Ref. \onlinecite{Padilla2005}). One should note that the Fermi velocity and the Fermi wave vector are in direct relation to the effective mass through the relation $m^{\star}=\hbar k_F/v_
F$, which 
allows us to determine the dynamical value of $m^{\star}$ (black inverted triangles) by using ARPES measurements of $k_F$ (taken from Ref. \onlinecite{Hashimoto2008}) and $v_F$ (taken from Ref. \onlinecite{Zhou2003}). Note the differences between the two experimental data sets for $m^{\star}$ corresponding to LSCO (blue dots and black inverted triangles). Nevertheless, one universal feature of the results presented in Fig. \ref{fig:35}b is clearly visible. Namely, the effective mass in the nodal direction is almost constant, and if we ignore the upper data set, the value is $m^{\star}\backsimeq 2 m_e$, which is in very good agreement with our theoretical results within the GWF solution for the $t$-$J$-$U$ model. This value corresponds to the maximal value of $m^{\star}(\delta)$ determined recently for YBCO\cite{Ramshaw2015}, where its distinct non-universal dependence has been observed. The question 
arises as to what extent the dome-like behavior shown in Ref. \onlinecite{Ramshaw2015} can be related to charge-density-wave evolution in strong applied field.

\subsection{Subsidiary characteristics}
We have discussed already that the gain in the hopping (kinetic) energy is one of the crucial features differentiating between real-space-pairing models and singling out the $t$-$J$-$U$ model as the one that leads to a better agreement with experiment. However, according to the CDMFT calculations\cite{Haule2007} also the $t$-$J$ model gives some similar results in this respect (although they have not compared their results quantitatively to experiment). The agreement shown by us (cf. Figs. \ref{fig:3}, \ref{fig:3_1}, \ref{fig:35}) is achieved only if the high-energy scale (with $U=22.6|t|\sim 7.9$ eV), which is about three times larger than the bare bandwidth ($W\approx 2.8$ eV), is included in the analysis. To illustrate the role of such a high-energy resonant level, located at $E_g\equiv U-W \simeq 3$ eV, we have plotted in Fig. \ref{fig:4}a the value of $\Delta_G$ and $d^2=\langle \hat{n}_{i\uparrow}\hat{n}_{i\downarrow} \rangle_G$, both as a function of $U$. The small value of $d^2\lesssim 10^{-2}$, 
speaks 
in favor of the interpretation that the states in the upper Hubbard subband may play minor but still relevant role of high-energy resonant states. This circumstance can be put in accord with the canonical approach based on the split Hubbard subbands which are reproduced within the three band model - the original model describing the Cu-O plane, as discussed in Appendix. In Fig. \ref{fig:4}b we show the doping dependence of the variational parameter $x$, which is introduced in the constraint (\ref{eq:constraint}), also for the case of the $t$-$J$-$U$ model. For the half-filled situation we obtain the value $x=-4$, which corresponds to no double occupancy, in spite of having a finite value of the Hubbard $U$. This particular value of $x$ results from the fact that for the case with $d^2=0$ we have $\lambda_d=0$ which leads directly to the relation $x=-1/(1-n_0)^2$ (cf. Section Model and Method as well as Ref. \onlinecite{Kaczmarczyk2014}). Hence, the $t$-$J$-$U$ both in the $U\rightarrow \infty$ limit and for 
the half-filled case with the finite $U$ leads to the 
results equivalent to the one obtained for the $t$-$J$ model. In Fig. \ref{fig:4}c we provide $\Delta E_{\mathrm{kin}}$ vs $\delta$ for the limiting case $J=0$ and $d^2\equiv 0$, which may 
be regarded as the $U\rightarrow \infty$ limit of both the Hubbard and the $t$-$J$ models. One can see that even though the general shape of the curve in 
Fig. \ref{fig:4}c is the same as that obtained for the $t$-$J$-$U$ model (cf. Fig. \ref{fig:3}c), the corresponding 
values of $\Delta E_{\textrm{kin}}$ are smaller by at least an order of magnitude. Therefore, only by combining the two factors, nonzero $d^2$ (finite $U$) and a relatively large value of $J$, one can bring the theory in the quantitative accord 
with experiment, at least within the Gutzwiller wave function solution. This required a generalization of the ideas of kinetic exchange as it comes out from a direct perturbation expansion of the Hubbard model\cite{Chao1976,Spalek2007}, as mentioned above. The fundamental question concerning the model is whether other methods of approach (VMC, DMFT and others) can be applied to it and confirm the presented here results obtained within the DE-GWF method. 
The affirmative answer to this question would constitute, in our view, a basis for comprehensive treatment of the pairing as applied to high temperature superconductors and other strongly correlated systems.

\section{Discussion and Outlook}
The results presented in Figs. \ref{fig:1} - \ref{fig:35} provide a consistent analysis for the same set of model parameters of the principal experimental properties of the cuprates within the combined concepts of real space pairing and strong interelectronic correlations. In carrying out our analysis we had to go beyond the renormalized mean field theory (even in its statistically consistent version, SGA\cite{Jedrak2011}), i.e., discuss the results within the full Gutzwiller wave function solution (GWF) to a relatively high order of the diagrammatic expansion. In particular, we have explained here the following ground-state characteristics: {\bf (i)} the doping ($\delta$) independence of the Fermi velocity $v_F$ in the nodal direction, {\bf (ii)} the kinetic energy gain in the SC phase $\Delta E_{\mathrm{kin}}$, one of the main non-BCS features, {\bf (iii)} the optimal doping value $\delta_{\mathrm{opt}}\approx 0.16-0.2$, {\bf (iv)} the upper critical concentration for 
disappearance of 
the HTS state, $\delta_{c2}\lesssim 0.4$, and {\bf (v)} 
the doping 
dependence of the Fermi wave vector, $k_F(\delta)$. Additionally, we have extracted the $\delta$ dependence of effective mass enhancement from the experimental data concerning $k_F(\delta)$ and $v_F(\delta)$ and have shown that $(m^{\star}/m_e)|_{E=E_F}$ agrees well with that obtained theoretically, as well as that determined from an independent experiments.

The $t$-$J$-$U$ model taken at the start requires a slight modification of our thinking about HTS as doped Mott insulators in 
terms of either the Hubbard or the original $t$-$J$ models \cite{Chao1976,Anderson1988,Spalek1988,Kotliar1988,Ruckenstein1987}. In this respect, one formal point of the model should be noted here. Namely, the antiferromagnetic exchange is quite strong and must be coming from the $d-d$ superexchange via the antibonding $2p_{\sigma}$ states due to oxygen, as stated many times earlier \cite{Ogata2008,Zhang1988,Zaanen1990,Eskes1993,Jefferson1992,Feiner1996,Avella2013}. In effect, as the fitting to the experimental data provides us a posteriori with the Hubbard interaction to the bare band-width ratio $U/W\simeq 2.5$, the electronic correlations in the effective narrow band may not be regarded as extremely strong, particularly for $\delta>0$, what results also in having a small but relevant double occupancy probability $d^2\lesssim 10^{-2}$. The presence of the Hubbard interaction term introduces in an explicit form the high energy scale $U\sim 8$ eV to the problem, what results in comparable values of the 
effective Coulomb energy $Ud^2\lesssim 80$ 
meV and the 
exchange energy $J\sim 120$ meV. Moreover, the 
kinetic energy in the PM 
state, is also of the same order, $4|t|\delta\backsimeq 140$ meV for $\delta=0.1$, constituting altogether a truly correlated state, in which all the three factors play a role. In connection with taking finite $U$ value it should be noted that the holon-doublon correction to the Gutzwiller wave-function have also been considered \cite{Ogata2008}. It would be interesting to see the connection between that extension and our approach. 

In our considerations we have disregarded the intersite Coulomb repulsion (the third term of Eq. \ref{eq:H_start})), as it does not influence much the quality of the comparison with the discussed here experimental data. Explicitly, the LSCO data displayed in Figs. \ref{fig:3}a and \ref{fig:3_1} can be also fitted with the set of parameters: $t=-0.35$eV, $t'=-0.25|t|$, $J=0.3|t|$, $U=22|t|$, $V=0.2|t|$, i.e. with $V\neq 0$. In effect, $V-J/4\approx 0.1|t|$ is small and can 
safely be disregarded here. However, 
the role of $V$ may become important when charge- and spin-density-wave states are included, but that requires a separate analysis\cite{Abram2016}.

We have not addressed at all the system thermodynamical properties. The extension to the temperature $T>0$ is indispensable as the next step. In this respect, particularly important is the question of the pseudogap appearance\cite{Huffner2008,Kordyuk2015}. It is intriguing to ask whether the pseudogap is partly connected with the evolution of our effective gap $\Delta^{\mathrm{eff}}$ in the antinodal direction or is it due to a different physical mechanism\cite{Rice2012}. A possible connection between the effective gap and the measured gap in the antinodal direction is supported by the intriguing coincidence that both of them increase with the decreasing doping\cite{Kaczmarczyk2014,Yoshida2009}. A similar behavior has already been obtained within SGA by taking the bare (not Gutzwiller-projected) value of the gap magnitude and fitting it to the experiment\cite{Edegger2007,Jedrak2011}. Furthermore, the appearance of other phases, such as spin- and charge-density-wave states on the 
superconducting phase diagram depicted in Fig. \ref{fig:2_1} should be treated 
separately, together with singling out the role of 
the intersite Coulomb interaction \cite{Lauglin2014,Abram2016}. 

The consistent scheme of analyzing concrete, though selected data for high-$T_C$ SC phase is not the last word by any means, also due to the following reasons. It would be interesting to compare the present results with those of other methods, which also go beyond the renormalized mean field theory. For example, an application of the plaquette or cluster dynamic mean-field theory\cite{Frantino2016, Haule2007} to the present model could be of principal importance as an independent checkout on the validity of higher order corrections to RMFT. Furthermore, an extension of our approach to the situation with nonzero applied magnetic field would provide additional physical properties (e.g., doping dependence of the penetration depth) for a further quantitative testing of the present approach. We should be able to see progress along these lines in the near future.

\section{Acknowledgements}

The authors are grateful for the financial support of the National Science Centre (NCN) through Grant MAESTRO, No. DEC-2012/04/A/ST3/00342. The discussions with Profs. Dirk van der Marel from Universit\'e de Geneve, Adam Kami\'nski from the Iowa State University and the AMES Lab., Iowa, and Alexander Kordyuk from the National Academy of Ukraine, were useful and enlightening.

\appendix
\section{Methodological discussion: physical significance of the extended model}
We would like to estimate the physical relevance of the considered here $t$-$J$-$U$ model. In the canonically transformed extended Hubbard model\cite{Anderson1988,Spalek2007,SpalekOles} the antiferromagnetic exchange interaction is of the form $J_{ij}=2t_{ij}^2/(U-V_{ij})$ and therefore no Hubbard extra term should appear\cite{Harris}, if we are in the strong-correlation limit $W<<U$ (not only $|t_{ij}|<<U$). Namely, the contribution to the $N$-particle wave function coming from double occupancies is of the order of $t/|U|$\cite{Harris}. Before discussing the application of that limit in real calculations, let us estimate its value by taking the standard microscopic-parameter: $U=8-10\;$eV, $t=-0.35\div-0.4\;$eV, and $t'=|t|/4$. When neglecting $V$ in above formula for $J_{ij}$ we obtain the value of $J\simeq 150\;$K at most. If the bare parameter $V$ is taken as $U/3$ (maximum), then the value of $J$ increases by $50\%$, which is still much lower than the typical value of measured $J\sim 0.13\;$eV$\approx 1.
5\cdot 10^3$K in the insulating phase\cite{Harris,Kastner} (note that here the values of $J$ are $1/2$ of those for 
the full 
exchange as there is no factor $1/2$ before the last term in (\ref{eq:H_start})). On the other hand, the bare bandwidth of the planar states is $W=8|t|\simeq2.8\;$eV$\sim 3\;$eV. Therefore, the $U/W$ ratio is in the interval $2.9-3.6\sim 3$, which is not in the asymptotic limit of being $\gg 1$. Hence, one may expect that the double occupancy probability is not exactly vanishing, particularly for $\delta>0$ as then the admixture of double occupancy to the single-particle state is of the order\cite{Harris} of $|t|/U\sim 0.04$. In our calculations (cf. Fig. \ref{fig:4}) $d\lesssim 10^{-2}$ for $\delta=0.1$ and $U/W=2.5$. Such a small value does not influence at all the spin magnitude in the Mott insulating state, since then $\langle \underline{S}^{2}\rangle=(3/4)(1-2d^2)\sim 1/2(1/2+1)+o(10^{-2})$\cite{Spalek1990} and the zero-point spin fluctuations are much more important. 

After mentioning the relevance of the Hubbard term, the basic question still remains as to what is the dominant contribution to $J$. As said earlier, this is due to the superexchange\cite{Zaanen1990,Eskes1993,Jefferson1992,Feiner1996,Avella2013} via $p$ orbitals with inclusion of the fact that HTS are charge transfer insulators with the corresponding gap $\Delta=\epsilon_p-\epsilon_d\simeq 3\;$eV and the $p-d$ hybridization magnitude $t_{pd}\simeq 1.3\;$eV, as well as the $p-d$ Coulomb interaction $U_{pd}\simeq1$ eV. In effect, the nearest neighbors superexchange can be estimated as\cite{Zaanen1990,Eskes1993,Jefferson1992,Avella2013} 
\begin{equation}
\begin{split}
 J&=\frac{2\;t_{pd}^4}{(\Delta_{pd}+U_{pd})^2}\bigg(\frac{1}{U_{d}}+\frac{1}{\Delta_{pd}+U_{pp}/2}\bigg)\\
 &\simeq 0.13\;\textrm{eV}=1560\;\textrm{K},
 \end{split}
\end{equation}
a value close to that determined experimentally\cite{Kastner}. This reasoning provides a direct support for the effective value of $J$ as not coming from the large-$U$ expansion of the Hubbard model\cite{Chao1976,Spalek1988,Spalek2007}.

One should note that the mechanism introduces also the Kondo-type coupling between the $p$ holes and $d$ electrons with the corresponding Kondo exchange integral 
\begin{equation}
 J_K=2\;t_{pd}^2\bigg(\frac{1}{\Delta_{pd}}+\frac{1}{{\Delta_{pd}+U}}\bigg)\simeq 2.5\;\textrm{eV}\gg J.
 \label{eq:Kondo}
\end{equation}
This coupling causes a bound configuration of the hole and $d$-electron of Cu$^{2+}$ ion composing the Zhang-Rice singlet\cite{Zhang1990,Zhang1988}. Also, the hopping amplitude for $d$ electron between the nearest neighboring sites $<i,j>$ can be estimated as
\begin{equation}
 t\sim\frac{t^2_{pd}}{(\Delta_{pd}+U_{pp}-U_{pd})^2}t_{pp}.
\end{equation}
Taking $t_{pp}=0.1\;$eV and $U_{pp}=3\;$eV, we obtain $t=-0.34\;$eV, also a quite reasonable value. In effect, we have $J/|t|=0.38$ which is a reasonable ratio in view of simplicity of our estimates. In such a reduction procedure to the one-band model the effective Hubbard interaction is $U\simeq U_{dd}-U_{pp}\simeq 7.5\;$eV, if we take the value $U_{dd}=10.5\;$eV for the original $d$ atomic states. In the fitting to experiment we have obtained a slightly larger value of $U=8\;$eV and $|t|=0.35\;$eV. This brief discussion summarizes the meaning of the starting Hamiltonian (\ref{eq:H_start})\cite{Zhang1990,Zhang1988,SpalekOles,Zaanen1990,Eskes1993,Jefferson1992,Feiner1996,Avella2013}.

The general 3-band model would include a direct single-particle hopping between the oxygen sites $\sim t_{pp}$. Under these circumstances, the Kondo-type coupling (\ref{eq:Kondo}) between the oxygen and copper sites must be also taken into account explicitly and we end up in the Emery-Reiter type of model\cite{Emery1988,Valkov} in the limit of localized $3d$ electrons. In this respect, we have selected here the most general one-band model of correlated $d$ electrons with the Zhang-Rice singlet idea implicitly assumed.

%\cite{Zaanen1990,Eskes1993,Jefferson1992,Feiner1996,Avella2013,Emery1988,Valkov}

\end{document}